\documentclass[letter]{aa} 
\usepackage{graphicx}
\usepackage{txfonts}
\begin{document}

\title{\ion{H}{i} column densities of $z > 2$ \emph{Swift} gamma-ray bursts}
 
\author{
P.~Jakobsson \inst{1,2,3}
\and
J.~P.~U.~Fynbo \inst{2}
\and
C.~Ledoux \inst{4}
\and
P.~Vreeswijk \inst{4,5}
\and
D.~A.~Kann \inst{6}
\and
J.~Hjorth \inst{2}
\and
R.~S.~Priddey \inst{1}
\and
N.~R.~Tanvir \inst{7}
\and
D.~Reichart \inst{8}
\and
J.~Gorosabel \inst{9}
\and
S.~Klose \inst{6}
\and
D.~Watson \inst{2}
\and
J.~Sollerman \inst{2}
\and
A.~S.~Fruchter \inst{10}
\and
A.~de~Ugarte~Postigo \inst{9}
\and
K.~Wiersema \inst{11}
\and
G.~Bj\"ornsson \inst{3}
\and
R.~Chapman \inst{1}
\and
C.~C.~Th\"one \inst{2}
\and
K.~Pedersen \inst{2}
\and
B.~L.~Jensen \inst{2}
}


\institute{
Centre for Astrophysics Research, University of Hertfordshire,
College Lane, Hatfield, Herts, AL10 9AB, UK
\and
Dark Cosmology Centre, Niels Bohr Institute, University of
Copenhagen, Juliane Maries Vej 30, 2100 Copenhagen, Denmark
\and
Science Institute, University of Iceland, Dunhaga 3, 
107 Reykjav\'{\i}k, Iceland
\and
European Southern Observatory, Alonso de C\'ordova 3107, Casilla 19001, 
Vitacura, Santiago, Chile
\and
Departamento de Astronom\'{\i}a, Universidad de Chile, Casilla 36-D, 
Santiago, Chile
\and
Th\"uringer Landessternwarte Tautenburg, Sternwarte 5, 
D-07778 Tautenburg, Germany
\and
Department of Physics and Astronomy, University of Leicester, 
Leicester LE1 7RH, UK
\and
Department of Physics and Astronomy, University of North Carolina at 
\mbox{Chapel Hill, Campus Box 3255, Chapel Hill, NC 27599, USA}
\and
Instituto de Astrof\'{\i}sica de Andaluc\'{\i}a (CSIC), 
Apartado de Correos 3004, 18080 Granada, Spain
\and
Space Telescope Science Institute, 3700 San Martin Drive, Baltimore, 
MD 21218, USA
\and
Astronomical Institute `Anton Pannekoek', University of Amsterdam, 
Kruislaan 403, 1098 SJ Amsterdam, The Netherlands
}

\date{Received 15 September 2006 / Accepted 10 October 2006}

 
  \abstract
  {Before the launch of the \emph{Swift} satellite, the majority of the
  gamma-ray burst (GRB) afterglows for which Ly$\alpha$ was redshifted
  into the observable spectrum showed evidence for a damped Ly$\alpha$ 
  absorber. This small sample indicated that GRBs explode either in
  galaxies, or regions within them, having high neutral hydrogen column 
  densities.}   
  {To increase the spectroscopic sample of GRBs with $z>2$ and hence 
  establish the $N(\ion{H}{i})$ distribution along GRB lines-of-sight.}
  {We have obtained six $z > 2$ GRB afterglow spectra and fitted the 
  Ly$\alpha$ absorption line in each case to determine $N(\ion{H}{i})$.
  This has been complemented with 12 other \emph{Swift} $N(\ion{H}{i})$ 
  values from the literature.}
  {We show that the peak of the GRB $N(\ion{H}{i})$ distribution is 
  qualitatively consistent with a model where GRBs originate in Galactic-like 
  molecular clouds. However, a systematic difference, in particular an 
  excess of low column-density systems compared to the predictions, indicates 
  that selection effects and conditions within the cloud (e.g. strong 
  ionization) influence the observed $N(\ion{H}{i})$ range. We also report 
  the discovery of Ly$\alpha$ emission from the GRB\,060714 host, 
  corresponding to a star-formation rate of approximately 
  0.8\,$M_{\odot}\,\textrm{yr}^{-1}$. Finally, we present accurate redshifts 
  of the six bursts: $z = 3.240 \pm 0.001$ (GRB\,050319), 
  \mbox{$z = 2.198 \pm 0.002$} (GRB\,050922C), $z = 3.221 \pm 0.001$ 
  (GRB\,060526), $z = 3.425 \pm 0.002$ (GRB\,060707), $z = 2.711 \pm 0.001$ 
  (GRB\,060714) and $z = 3.686 \pm 0.002$ (GRB\,060906).}
  {}


\maketitle


\section{Introduction}
In just 18 months, the \emph{Swift} satellite (Gehrels et al. \cite{swift}) 
has already contributed considerably to the progress of gamma-ray burst 
(GRB) science. It detects roughly two GRBs/week and transmits their 
accurate localizations (error radius frequently less than 5\arcsec) to the 
ground within minutes for rapid follow-up observations. The GRB redshift 
distribution found by \emph{Swift} is very different from that of the 
pre-\emph{Swift} sample, being skewed to much higher redshifts 
(Berger et al. \cite{bergerSwift}; Jakobsson et al. 
\cite{palli_swift,palli_swift2}; Daigne et al. \cite{daigne}; Le \& Dermer
\cite{le}). Approximately 70$\%$ of the \emph{Swift} bursts are found to be 
located at $z > 2$, while the corresponding fraction was only 20$\%$ for 
pre-\emph{Swift} bursts.
\par
It is thus much more common for the Ly$\alpha$ line in \emph{Swift} bursts
to be redshifted redward of the atmospheric cutoff in optical spectra.
This provides us with the opportunity to investigate the GRB host neutral 
hydrogen column density distribution. In particular we can compare it to 
the damped Ly$\alpha$ absorbers (DLAs) seen in absorption against QSO
spectra (see Wolfe et al. \cite{wolfe} for a recent review), defined as 
displaying $N(\ion{H}{i}) \geq 2 \times 10^{20}$\,cm$^{-2}$ 
$(\log N(\ion{H}{i}) \geq 20.3)$. Most of the neutral gas in the Universe 
in the redshift interval $0 < z < 5$ is in DLAs, providing the fuel for 
star formation at these epochs. Given that long-duration GRBs are known to 
have massive stellar progenitors (e.g. Hjorth et al. \cite{jens}; Malesani 
et al. \cite{daniele}; Fruchter et al. \cite{andy}), GRB-DLAs provide 
valuable information on the sites of active star formation in the 
high-redshift Universe. In the pre-\emph{Swift} era, only seven GRB 
$N(\ion{H}{i})$ column density measurements were obtained, with six being 
classified as DLAs (Vreeswijk et al. \cite{paul} and references therein).
\par
In this Letter we present optical spectroscopy of six GRBs, focusing
on the measurement of the $N(\ion{H}{i})$ column density. We also report 
the detection of Ly$\alpha$ emission from one of the bursts (GRB\,060714). 
We then discuss how the observed $N(\ion{H}{i})$ column density distribution 
from \emph{Swift} bursts can be understood in terms of the GRB environment.

\section{Observations}
GRBs\,050319, 050922C, 060526, 060707, 060714 and 060906 are all 
long-duration bursts detected by the \emph{Swift} satellite. Each burst 
was localized by the Burst Alert Telescope, each position refined
by the X-Ray Telescope and a subsequent optical afterglow (OA) was detected 
in every case (Rykoff et al. \cite{rykoff319,rykoff922C}; Campana et al. 
\cite{campana}; de Ugarte Postigo \cite{antonio}; Krimm et al. \cite{krimm}; 
Cenko et al. \cite{cenko906}).
\par
Using the Nordic Optical Telescope (NOT), we obtained spectra of the
OA of GRBs\,050319 and 050922C (Fynbo et al. \cite{fynbo_319}; Jakobsson
et al. \cite{palli_922}). The data were acquired with the ALFOSC instrument
with a 1\farcs3 wide slit and a grism with a wavelength coverage from
3700--9100\,\AA. The FORS1 instrument on the Very Large Telescope (VLT)
was used to obtain OA spectra of GRBs\,060526, 060707, 060714 and 060906
(Th\"one et al. in prep; Jakobsson et al. \cite{palli_707, palli_714}; 
Vreeswijk et al. \cite{paul906}). A 1\farcs0 wide slit was used and grisms 
with a wavelength coverage from 4700--7100\,\AA\ (GRB\,060526) and 
3600--9000\,\AA. The details of our observations are given in 
Table~\ref{obs.tab}.
\section{Results}
The spectrum of each OA displays a strong absorption line with usually 
clear damped wings. Blueward of this line, the flux drops substantially
and exhibits the signature of the Ly$\alpha$ forest. Associating the line 
with Ly$\alpha$, the rest of each spectrum was searched for additional 
features ($>$3$\sigma$) at the corresponding approximate redshift. The
results are presented in Table~\ref{res.tab}.
\par
There is additional strong absorption in GRB\,050922C, \mbox{corresponding} to 
Ly$\alpha$ at $z \approx 2.07$. At this redshift we identify four other 
features (\ion{O}{i} 1302/\ion{Si}{ii} 1304, \ion{C}{ii} 1334/\ion{C}{ii}* 
1335, \ion{Si}{ii} 1526 and \ion{C}{iv} 1548,1550) with an average redshift 
of $z = 2.075 \pm 0.003$. 
\par
\begin{table}
\caption[]{A log of the follow-up spectroscopic observations for the
six bursts presented in the paper. $\Delta t$ is the time from the onset 
of the burst.}
\label{obs.tab}
\centering
\setlength{\arrayrulewidth}{0.8pt}   
\begin{tabular}{@{}llrcc@{}}
\hline
\hline
\vspace{-2 mm} \\
GRB & Tel/instrument/grism & Exposure & $\Delta t$ & Spectral \\
    &                & time [s] & [days]     & res. [\AA] \\
\hline
\vspace{-2 mm} \\
050319  & NOT/ALFOSC/\#4 & $3 \times 2400$ & 1.50 & 7 \\
050922C & NOT/ALFOSC/\#4 & $2400$          & 0.05 & 5 \\
060526  & VLT/FORS1/600V & $900 + 1800$    & 0.44 & 5 \\
060707  & VLT/FORS1/300V & $3 \times 1800$ & 1.44 & 10 \hspace*{1.0 mm}\\
060714  & VLT/FORS1/300V & $3 \times 1800$ & 0.50 & 8 \\
060906  & VLT/FORS1/300V & $600$           & 0.05 & 8 \\
\hline
\end{tabular}
\end{table}
\begin{figure}
\centering
\resizebox{\hsize}{!}{\includegraphics[bb=114 -120 471 798,clip]{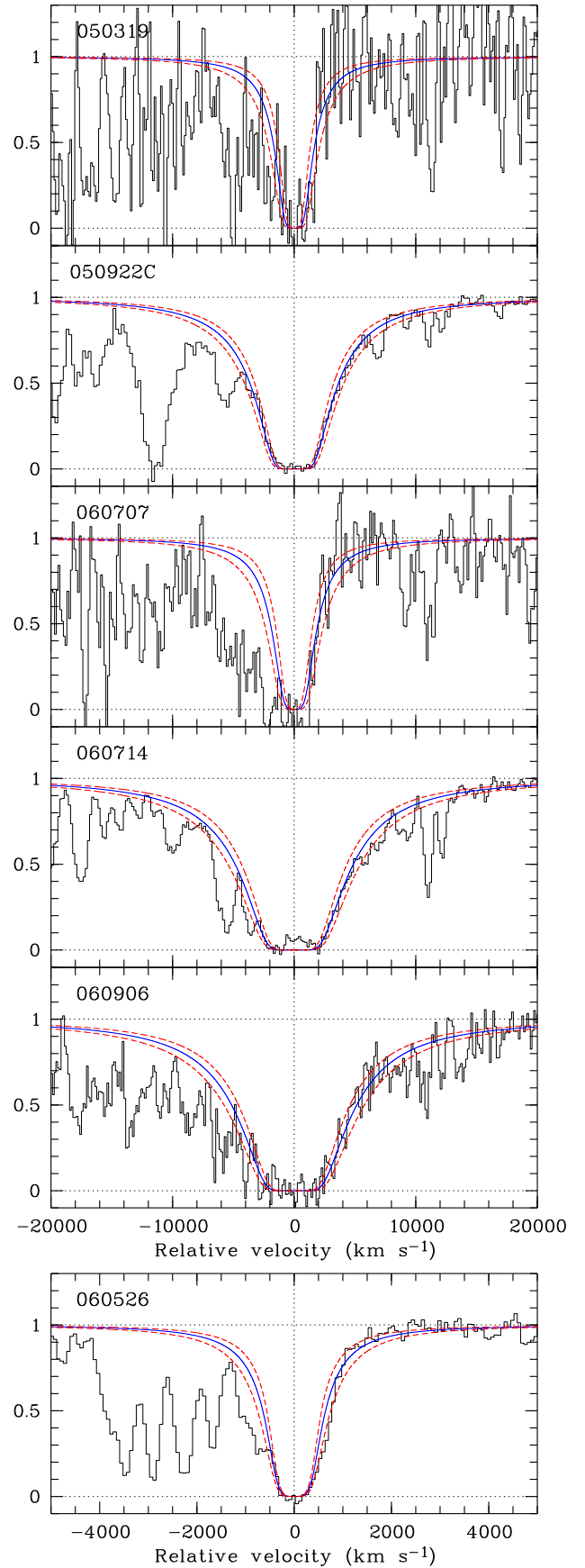}}
\caption{The normalized OA spectra centred on the Ly$\alpha$
         absorption lines at the GRB host galaxy redshifts. A neutral 
         hydrogen column density fit to the damped Ly$\alpha$ line is 
         shown with a solid line in each panel, while the 1$\sigma$ 
         errors are shown with dashed lines.}
\label{dla.fig}
\end{figure}
In Fig~\ref{dla.fig} we plot the normalized OA spectral region around
Ly$\alpha$ for each burst. Overplotted is a fit to the Ly$\alpha$
absorption line yielding values reported in Table~\ref{res.tab}. Apart 
from GRB\,060526, these values are well above the DLA definition. The 
redshifts deduced from the metal lines were found to be perfectly consistent 
with the DLA line fits for GRBs\,050922C and 060714. For the other four 
spectra, we note that slightly lower redshifts were used in order to provide 
the best fit to both the core and the red damped wing of the Ly$\alpha$ 
profiles. However, adopting the metal line redshifts would make the 
$N(\ion{H}{i})$ only slightly smaller and within the reported 1$\sigma$ 
uncertainties.
\begin{table*}
\caption[]{The redshift, neutral hydrogen column density and absorption
lines identified for each afterglow spectrum. The number in square brackets 
after the redshift indicates the number of lines used to calculate it.}
\label{res.tab}
\centering
\setlength{\arrayrulewidth}{0.8pt}   
\begin{tabular}{@{}llcl@{}}
\hline
\hline
\vspace{-2 mm} \\
GRB & \hspace*{7.5 mm} $z$ & $\log N(\ion{H}{i})$ & Absorption features \\
\hline
\vspace{-2 mm} \\
050319  & $3.240 \pm 0.001$ [3] & $20.90 \pm 0.20$ & Ly$\alpha$, 
\ion{Si}{ii} 1260, \ion{O}{i} 1302, \ion{C}{ii} 1334/\ion{C}{ii}* 1335, 
\ion{C}{iv} 1548,1550 \\
050922C & $2.198 \pm 0.002$ [8] & $21.55 \pm 0.10$ & Ly$\alpha$, 
\ion{Si}{ii} 1260, \ion{Si}{ii}* 1264, \ion{O}{i} 1302/\ion{Si}{ii} 1304, 
\ion{C}{ii} 1334/\ion{C}{ii}* 1335, \ion{Si}{iv} 1393,1402, \\ 
& & & \ion{Si}{ii} 1526, \ion{Si}{ii}* 1533, \ion{C}{iv} 1548,1550,
\ion{Fe}{ii} 1608, \ion{Al}{ii} 1670 \\
060526  & $3.221 \pm 0.001$ [12] & $20.00 \pm 0.15$ & \ion{Si}{ii} 1190,1193,
\ion{Si}{iii} 1206, Ly$\alpha$, \ion{Si}{ii} 1260, \ion{O}{i} 1302, 
\ion{Si}{ii} 1304, \ion{C}{ii} 1334, \\
& & & \ion{C}{ii}* 1335, \ion{Si}{ii} 1526, \ion{Si}{ii}* 1533, 
\ion{Fe}{ii} 1608, \ion{Al}{ii} 1670 \\
060707  & $3.425 \pm 0.002$ [8] & $21.00 \pm 0.20$ & Ly$\beta$, 
\ion{C}{ii} 1036/\ion{C}{ii}* 1037, \ion{N}{ii} 1083, \ion{Si}{iii} 1206, 
Ly$\alpha$, \ion{Si}{ii} 1260, \ion{O}{i} 1302/\ion{Si}{ii} 1304, \\
& & & \ion{C}{ii} 1334/\ion{C}{ii}* 1335, \ion{Si}{ii} 1526, 
\ion{C}{iv} 1548,1550, \ion{Al}{ii} 1670 \\
060714  & $2.711 \pm 0.001$ [36] & $21.80 \pm 0.10$ & Ly$\gamma$, 
\ion{C}{iii} 977, \ion{Si}{ii} 989, Ly$\beta$, \ion{O}{vi} 1031,1037, 
\ion{C}{ii} 1036, \ion{C}{ii}* 1037, \ion{N}{ii} 1083, \ion{Fe}{ii} 1144, \\
& & & \ion{Si}{ii} 1190,1193, Ly$\alpha$, \ion{S}{ii} 1250,1253, 
\ion{Si}{ii} 1260, \ion{Si}{ii}* 1264, \ion{O}{i} 1302/\ion{Si}{ii} 1304, 
\ion{Si}{ii}* 1309, \\
& & & \ion{C}{ii} 1334/\ion{C}{ii}* 1335, \ion{Ni}{ii} 1317,1370, 
\ion{Si}{iv} 1393,1402, \ion{Si}{ii} 1526, \ion{Si}{ii}* 1533, 
\ion{C}{iv} 1548,1550, \\
& & & \ion{Fe}{ii} 1608, \ion{Al}{ii} 1670, \ion{Ni}{ii} 1709,1741, 
\ion{Si}{ii} 1808, \ion{Al}{iii} 1854,1862, \ion{Zn}{ii} 2026,2062, \\
& & & \ion{Fe}{ii} 2260,2344,2374,2382, 
\ion{Fe}{ii}** 2328,2349, \ion{Fe}{ii}*** 2359, \ion{Fe}{ii}* 2365 \\
060906  & $3.686 \pm 0.002$ [5] & $21.85 \pm 0.10$ & Ly$\beta$, 
\ion{C}{ii} 1036/\ion{C}{ii}* 1037, Ly$\alpha$, \ion{Si}{ii} 1260, 
\ion{O}{i} 1302/\ion{Si}{ii} 1304, \\
& & & \ion{C}{ii} 1334/\ion{C}{ii}* 1335, \ion{Si}{iv} 1393,1402, 
\ion{Si}{ii} 1526 \\
\hline
\end{tabular}
\end{table*}
\par
\begin{figure}
\centering
\resizebox{\hsize}{!}{\includegraphics{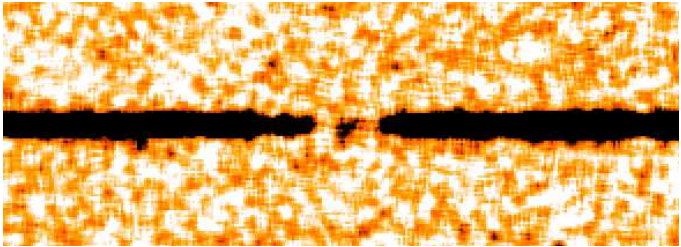}}
\caption{The two-dimensional GRB\,060714 OA spectrum centred on the 
         Ly$\alpha$ absorption line. Ly$\alpha$ in emission is clearly
         extended and centred in the trough.}
\label{ly-emission.fig}
\end{figure}

There is clearly a Ly$\alpha$ emission in the centre of the GRB\,060714
trough (Fig~\ref{ly-emission.fig}, see also Fig~\ref{dla.fig}), with an
approximate flux of $1.3 \times 10^{-17}$\,erg\,s$^{-1}$\,cm$^{-2}$.
In our assumed cosmology, ($\Omega_{\mathrm{m}}$, $\Omega_{\Lambda}$, $h$) =
(0.3, 0.7, 0.7), this corresponds to a luminosity of
$7.9 \times 10^{41}$\,erg\,s$^{-1}$. We can use this result to derive the
star-formation rate (SFR), assuming that a Ly$\alpha$ luminosity of
$10^{42}$\,erg\,s$^{-1}$ corresponds to a SFR of 
$1\,M_{\odot}\,\textrm{yr}^{-1}$ (Kennicutt \cite{kennicutt}; Cowie \& Hu 
\cite{hu}). The Ly$\alpha$ SFR in the GRB\,060714 host is thus 
$\sim$0.8\,$M_{\odot}\,\textrm{yr}^{-1}$, which is within the range of values 
found for other GRB hosts (e.g. Fynbo et al. \cite{fynbo_ly,fy}; 
Jakobsson et al. \cite{palliMNRAS}). We note that this value 
has not been corrected for host extinction, and is therefore a strict lower 
limit to the actual SFR. Fynbo et al. (\cite{fy}) find that Ly$\alpha$ 
emission is much more frequent among pre-\emph{Swift} GRB host galaxies than 
among the Lyman-break galaxies at similar redshifts and that a low metallicity 
preference for GRBs could be the explanation (see also Vink \& de Koter 
\cite{vink}; Fruchter et al. \cite{andy}; Priddey et al. \cite{rob}). 
We are currently undertaking a survey of GRB host galaxies that will 
constrain the fraction of Ly$\alpha$ emitters among \emph{Swift} GRB hosts.
\par
Although there are many metal lines observed in the GRB spectra, most of
them are saturated. However, we have identified a couple of lines,
corresponding to elements that are known to deplete negligibly on dust. 
They fall well outside the Ly$\alpha$ forest and are thus probably unblended. 
In addition, these transition lines are very weak, implying they are the
least saturated ones: \ion{Si}{ii} 1808 and \ion{Zn}{ii} 2026 in GRB\,060714. 
In the optically thin limit approximation, their equivalent widths 
correspond to the following metallicity limits: $\textrm{[Si/H]} \ge -1.35$ 
and $\textrm{[Zn/H]} \ge -1.0$. 

\section{Discussion}

\begin{table}
\caption[]{A list of all \emph{Swift} GRBs with $z > 2$ known to date
(1 October 2006). $N(\ion{H}{i})$ is derived from optical spectroscopy.
References are given in order for the redshift and the $\ion{H}{i}$ column 
density (information is not currently available for bursts marked with ---). 
(1) This work; 
(2) Watson et al. (\cite{watson06}); 
(3) Berger et al. (\cite{berger505}); 
(4) Berger et al. (\cite{berger603}); 
(5) Starling et al. (\cite{starling05}); 
(6) Ledoux et al. (\cite{ledoux05}); 
(7) Ledoux et al. (in prep); 
(8) Kawai et al. (\cite{kawai06}); 
(9) Foley et al. (\cite{foley05}); 
(10) Quimby et al. (\cite{quimby05}); 
(11) Piranomonte et al. (\cite{pira05}); 
(12) Prochaska et al. (\cite{prochaska06}); 
(13) J. X. Prochaska (private communication); 
(14) Fynbo et al. (\cite{fynbo06}); 
(15) Cucchiara et al. (\cite{cucchiara06}); 
(16) Cucchiara et al. (\cite{cucchiara_poster}); 
(17) Berger et al. (\cite{berger223}); 
(18) Price (\cite{price06}); 
(19) Cenko et al. (\cite{cenko06}); 
(20) Ferrero et al. (\cite{ferrero}); 
(21) Ledoux et al. (\cite{ledoux06}); 
(22) Smette et al. (in prep);
(23) Rol et al. (\cite{evert});
(24) D'Elia et al. (\cite{delia});
(25) Fugazza et al. (in prep); 
(26) Fynbo et al. (\cite{5.6}).}

\label{sample.tab}
\centering
\setlength{\arrayrulewidth}{0.8pt}   
\begin{tabular}{@{}llcl@{}}
\hline
\hline
\vspace{-2 mm} \\
GRB & \hspace{1.6 mm} $z$ & $\log N(\ion{H}{i})$ & References \\
\hline
\vspace{-2 mm} \\
050319  & 3.24 & $20.9 \pm 0.2$ & (1) (1) \\
050401  & 2.90 & $22.6 \pm 0.3$ & (2) (2) \\
050505  & 4.27 & $22.1 \pm 0.1$ & (3) (3) \\
050603  & 2.82 & ---  & (4) \\
050730  & 3.97 & $22.1 \pm 0.1$ & (5) (5) \\
050820A & 2.61 & $21.1 \pm 0.1$ & (6) (7) \\
050904  & 6.30 & $21.3$         & (8) (8) \\
050908  & 3.34 & $19.2$         & (9) (9) \\
050922C & 2.20 & $21.6 \pm 0.1$ & (1) (1) \\
051109  & 2.35 & ---  & (10) \\
060115  & 3.53 & ---  & (11) \\
060124  & 2.30 & $<$20.3 \hspace*{1.2 mm} & (12) (13) \\
060206  & 4.05 & $20.9 \pm 0.1$ & (14) (14) \\
060210  & 3.91 & $21.7 \pm 0.2$ & (15) (16) \\
060223  & 4.41 & ---  & (17) \\
060510B & 4.94 & ---  & (18) \\
060522  & 5.11 & $20.5 \pm 0.5$ & (19) (19) \\
060526  & 3.22 & $20.0 \pm 0.2$ & (1) (1) \\
060605  & 3.71 & ---  & (20) \\
060607  & 3.08 & $<$19.5 \hspace*{1.2 mm} & (21) (22) \\
060707  & 3.43 & $21.0 \pm 0.2$ & (1) (1) \\
060714  & 2.71 & $21.8 \pm 0.1$ & (1) (1) \\
060906  & 3.69 & $21.9 \pm 0.1$ & (1) (1) \\
060908  & 2.43 & ---  & (23) \\
060926  & 3.21 & $22.7 \pm 0.1$ & (24) (25) \\
060927  & 5.6  & ---  & (26) \\
\hline
\end{tabular}
\end{table}

\begin{figure}
\centering
\resizebox{\hsize}{!}{\includegraphics[bb=16 0 500 663,clip]{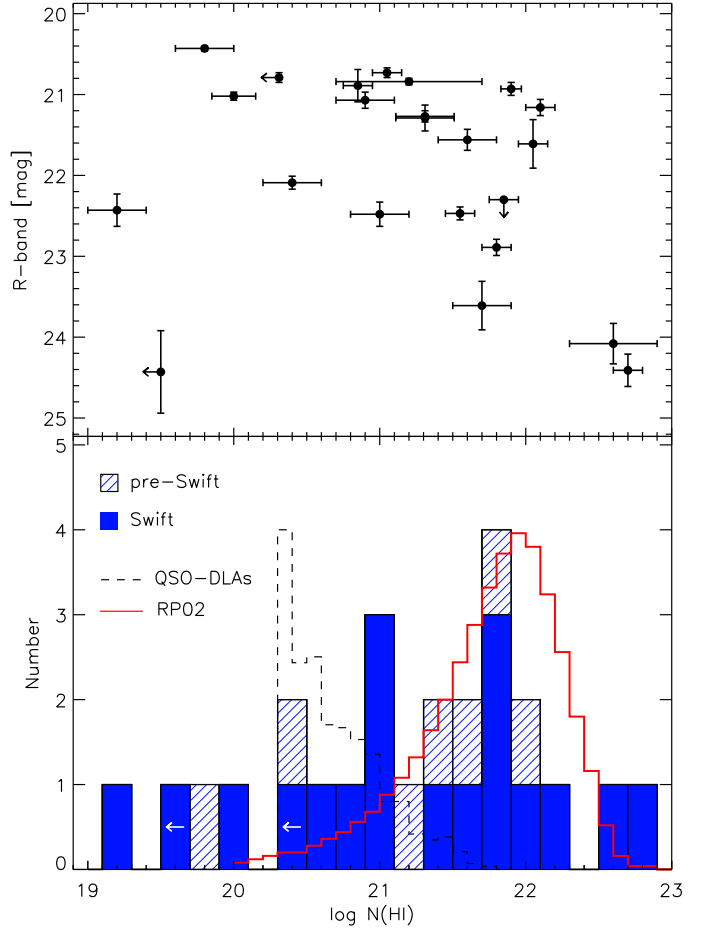}}
\caption{\emph{Bottom:} Histogram of the \ion{H}{i} column density
measured in GRBs for which the redshift was large enough to detect Ly$\alpha$. 
Bursts detected by \emph{Swift} are filled while the pre-\emph{Swift} sample 
is hashed (Vreeswijk et al. \cite{paul}). The current range of known GRB
\ion{H}{i} systems is $19.2 \leq \log N(\ion{H}{i}) \leq 22.7$. For 
comparison, the QSO-DLAs (Prochaska et al. \cite{QSO-DLA}) are overplotted
as the dashed histogram (normalized with respect to the GRB observations). 
The solid, non-filled histogram shows the GRB $N(\ion{H}{i})$ prediction by 
RP02. \emph{Top:} The corresponding afterglow $R$-band magnitudes, corrected 
for Galactic extinction (Schlegel et al. \cite{schlegel}), interpolated to a 
common epoch of 12\,hr (in the source restframe) and redshifted to $z = 3$. 
GRBs\,060210 and 060522 are omitted due to the lack of information on their 
spectral energy distribution; hence corrections could not be made to account 
for Ly$\alpha$ absorption. Where the uncertainty on the \ion{H}{i} column is 
not reported in the literature, a value of 0.2 has been plotted.}
\label{histo.fig}
\end{figure}

In Table~\ref{sample.tab} we have compiled all available $\ion{H}{i}$ 
column density measurements for $z > 2$ \emph{Swift} bursts (currently
displaying a median redshift of 3.4). The bottom panel of 
Fig.~\ref{histo.fig} shows a comparison between these $\ion{H}{i}$ 
column densities and those of the pre-\emph{Swift} sample (Vreeswijk et al. 
\cite{paul}). The former currently has a median $\log N(\ion{H}{i})$ of 
21.6, a bit higher than the latter (21.3). However, the small size of the 
pre-\emph{Swift} sample does not allow us to reject, with any degree of 
confidence, the null hypothesis that the two samples are drawn from the same 
distribution, i.e. a two-sample Kolmogorov-Smirnov (KS) test results in a 
90\% significance. The fraction of DLAs is also similar in both samples 
($\sim$80$\%$), supporting the conclusion that GRB absorption systems show 
exceptionally high column densities of gas when compared to DLA systems 
observed in the lines-of-sight to QSOs (Vreeswijk et al. \cite{paul}). 
Indeed, a two-sample KS test indicates there is a less than $10^{-7}$ 
probability that GRB-DLAs and QSO-DLAs (Prochaska et al. \cite{QSO-DLA}) are 
drawn from the same population. This result presumably reflects the fact 
that GRBs occur in star-forming regions within their hosts, whilst QSOs 
select more random lines-of-sight through intervening galaxies. 
\par
In the bottom panel of Fig.~\ref{histo.fig} we also compare the observed
GRB $N(\ion{H}{i})$ distribution to the expected column density distribution
for bursts in Galactic-like molecular clouds (Reichart \& Price 
\cite{reichart}, hereafter RP02). The model (solid, non-filled histogram)
is mass-weighted and corrected for geometrical effects, i.e. the clouds being 
centrally condensed, and the GRB location within a cloud (not behind it). 
Qualitatively, especially in terms of the location of the peak of the 
distribution, the match is relatively good. However, a KS test shows that the 
model and observed $N(\ion{H}{i})$ distributions are inconsistent with being 
drawn from the same population at the 3$\sigma$ level, with a clear 
overabundance of low-$N(\ion{H}{i})$ detections compared to the model. A 
possibility is that GRB host galaxies tend to have lower $N(\ion{H}{i})$ 
clouds than the Milky Way (MW) and/or GRBs are more likely to occur in 
lower $N(\ion{H}{i})$ clouds. Some evidence for the former exists 
(Rosolowsky \cite{MCs}), with the mass distribution of clouds in the LMC 
(which is more similar to GRB hosts than the MW) having a marginally steeper 
distribution than the inner disk of the MW.
\par
A few other effects can give rise to the excessive detections of 
$\log N(\ion{H}{i}) \leq 21.0$. The GRB progenitor and nearby massive stars 
may (partially) ionize their local environments; there is clearly a trend 
for GRBs with the smallest $N(\ion{H}{i})$ to exhibit very weak 
low-ionization lines (e.g. \ion{Si}{ii} and \ion{C}{ii}) while displaying 
strong absorption of e.g. \ion{Si}{iv} and \ion{C}{iv} (GRB\,021004: 
M\o ller et al. \cite{palle}; GRB\,050908: Prochaska et al. 
\cite{x_908}; GRB\,060124: Prochaska et al. \cite{prochaska06}). This may 
support the high-ionization scenario for the low-$N(\ion{H}{i})$ systems, 
although implicit is the assumption that low- and high-ionization species 
trace the same phase. An alternative scenario is that some GRBs are formed 
by massive runaway stars (e.g. Hammer et al. \cite{hammer}); these would
explode in regions of relatively low $N(\ion{H}{i})$. Finally, it has
been suggested (e.g. White et al. \cite{white}; Sugitani et al. 
\cite{sugitani}; Miao et al. \cite{miao}) that a significant part of star 
formation in molecular clouds takes place at their outer edges. The 
triggering of this star formation also blows the cloud open, resulting in 
these stars having (largely) unimpeded view outside the cloud and presumably 
a low $N(\ion{H}{i})$ column.
\par
The drop-off in detections above $\log N(\ion{H}{i}) \geq 22.0$ could be 
due to selection effects, i.e. the sample might be incomplete due
to high extinction. GRBs most likely do not burn through all of the dust
in their molecular clouds (e.g. fig.~4 in RP02), so some high-$N(\ion{H}{i})$ 
systems are likely dimmed by dust and consequently missed in the sample in 
Table~\ref{sample.tab} (see also Savaglio et al. \cite{sandra}; Vergani et al. 
\cite{vergani}). In the top panel of Fig.~\ref{histo.fig} we have plotted 
the corresponding afterglow $R$-band magnitudes (UV restframe) interpolated 
to a restframe epoch of 12\,hr and redshifted to \mbox{$z = 3$}. The majority 
of the bursts have a well-sampled light curve, allowing us to determine the 
magnitudes via spline interpolation (see appendix A in Kann et al. \cite{alex} 
for a detailed description of the method used). For the most recent bursts 
with a relatively poor data sampling, we have applied extrapolation and 
adopted large conservative error bars. There is only a tentative evidence of 
an anti-correlation between the OA flux and higher $N(\ion{H}{i})$ at the 
2$\sigma$ level as derived from Spearman's rank correlation. It should be 
noted that intrinsic GRB $R$-band afterglow light curves show evidence for 
a bimodal luminosity distribution (Kann et al. \cite{alex}; Liang \& Zhang 
\cite{liang}; Nardini et al. \cite{nardini}), that increases the scatter in 
the top panel of Fig.~\ref{histo.fig}. Excluding the upper limits, there is 
indeed an indication of bimodality in the data.
\par
\par
\par
The characteristics of the dust-to-gas ratio in the clouds are discernible
from the range of more than three orders of magnitude in the \ion{H}{i} 
column densities (lower panel of Fig.~\ref{histo.fig}), suggesting a 
similar range in the dust column (e.g. Predehl \& Schmitt \cite{predehl}). 
If dust were entirely responsible for all of the observed variation in 
the $R$-band (about five magnitudes; upper panel of Fig.~\ref{histo.fig}), 
the $A_R$ range would, at most, be of that order, limiting it to a maximum 
of $A_R \approx 5$\,mag. At the Galactic dust-to-gas ratio, the observed 
$A_R$ (UV restframe) corresponding to the maximum $N(\ion{H}{i})$ value is 
around 80 magnitudes (e.g. Pei \cite{pei}; Diplas \& Savage \cite{diplas}), 
implying that the $A_R/N(\ion{H}{i})$ ratio in these absorbers must be at 
most 7\% of the Galactic value (see also Hjorth et al. \cite{jens_124}). If 
some of the variation in the flux is due to other factors, the ratio drops 
even further.
\par
Given that it is thought that most of the star formation in the Universe
occurs in molecular clouds, it seems logical that long-duration GRBs
originate in such regions (Fig.~\ref{histo.fig}). However, there is a 
systematic difference between the model and observations that presumably 
can be clarified as a result of a combination of selection effects (dust 
extinction) and intense ionization in the GRB environment. Combining the
$N(\ion{H}{i})$ values derived from optical spectroscopy with the metal
column densities from soft X-ray absorption would conceivably shed further 
light on these issues. A detailed analysis of this is presented in a
subsequent paper (Watson et al. \cite{darach_NH}).

\begin{acknowledgements} We thank the referee, Evert Rol, for excellent
comments, and Jason X. Prochaska for providing us with the SDSS QSO-DLA 
data. PJ acknowledges PPARC for support, while NRT thanks PPARC for support 
through a Senior Research Fellowship. DAK and SK acknowledge support by 
DFG grant Kl 766/13-2. RSP and RC thank the University of Hertfordshire for 
a Research Fellowship and Studentship, respectively. The research of JG is 
supported by the Spanish Ministry of Science and Education through programmes 
ESP2002-04124-C03-01 and AYA2004-01515. GB acknowledges support from a 
special grant from the Icelandic Research Council. The Dark Cosmology Centre 
is funded by the Danish National Research Foundation. Based on observations 
made with the Nordic Optical Telescope, operated on the island of La Palma 
jointly by Denmark, Finland, Iceland, Norway, and Sweden, in the Spanish 
Observatorio del Roque de los Muchachos of the Instituto de Astrofisica de 
Canarias. Based on observations made with ESO Telescopes at the Paranal
Observatory under programme ID 077.D-0661(A+C).
\end{acknowledgements}

\end{document}